\documentclass[11pt]{article}
\usepackage[centertags]{amsmath}
\usepackage{amsfonts}
\usepackage{amssymb}
\usepackage{amsthm}
\usepackage{newlfont}
\usepackage{graphicx}
\usepackage{multirow}
\usepackage{rotating}
\usepackage{epstopdf}

\newtheorem{theorem}{Theorem}

\begin{document}
\thispagestyle{empty}

%
%

\title{ An efficient method to solve the mathematical model of HIV infection for CD8$^+$ T-cells   }
\author{ Samad Noeiaghdam$^{a,}$\footnote{Corresponding Author, E-mail addresses: s.noeiaghdam.sci@iauctb.ac.ir;
samadnoeiaghdam@gmail.com}~~~and~~~ Emran Khoshrouye Ghiasi$^b$
 }
  \date{}
 \maketitle

\begin{center}
\scriptsize{ $^a$Department of Mathematics, Central Tehran Branch, Islamic Azad University, Tehran, Iran. \\
$^b$Young Researchers and Elite Club, Mashhad Branch, Islamic Azad
University, Mashhad, Iran.}
\end{center}
\begin{abstract}
In this paper, the mathematical model of HIV infection for CD8$^+$
T-cells is illustrated.  The homotopy analysis method and the
Laplace transformations are combined for solving this model. Also,
the convergence theorem is proved to demonstrate the abilities of
presented method for solving non-linear mathematical models. The
numerical results for $N=5, 10$ are presented. Several
$\hbar$-curves are plotted to show the convergence regions of
solutions. The plots of residual error functions indicate the
precision of presented method.

 \vspace{.5cm}{\it keywords: Homotopy analysis method, Laplace transformations, Mathematical model of HIV infection, CD8$^+$ T-cells. }
\end{abstract}
\section{Introduction}

Human Immunodeficiency Virus (HIV) is one of the most dangerous
viruses in the world that leads to Acquired Immunodeficiency
Syndrome (AIDS). This virous involves the ribonucleic acid (RNA)
instead of the deoxyribonucleic acid (DNA) and finally the HIV
mechanism can be completed during 10-15 years \cite{hiv1}. In 1980,
the first case of HIV infection was reported. According to the
recent enumeration, more than 35 million people have been died by
HIV virous and more than 37 million people carry this virous in
their body and they are living as a menace on the world. Also, they
can transmit this threat by having unprotected sex, forwarding from
mother to child and other ways \cite{hiv2,hiv10,hiv3,hiv4}.

In last decades, many mathematical models have been presented to
identify the behavior of natural and artificial phenomena such as
mathematical model of HIV infection \cite{hiv6,man17,hiv5}, model of
Malaria viruses \cite{mal}, model of computer viruses
\cite{man14,man15,man18} and many other models \cite{omodel}. Also,
these models have been solved by many numerical or semi-analytical
methods.

The HAM is among of the semi-analytical methods which has been
presented by  Liao \cite{ham1,ham2,ham3,ham4,ham5}. In this method,
we have an operator, parameters and functions that we have freedom
to choose them. Selecting prepare parameters can lead to find the
solution of problem faster and more accurate than other
semi-analytical methods. In last decade, many authors applied the
HAM for solving mathematical and bio-mathematical problems such as
model of computer viruses \cite{man14}, model of HIV infection for
CD4$^+$ T-cells \cite{man17}, ill-posed problems \cite{man3} and
others \cite{em1,em2,em3,em4,em5,em6,em7}. Moreover, recently in
\cite{man7} we applied the CESTAC method \cite{man5,man11,man12} and
the CADNA library \cite{man16,man13} based on the stochastic
arithmetic to find the optimal step, the optimal error and the
optimal value of convergence control parameter of the HAM.

In some schemes, by combining the HAM by other methods or operators
we can construct new methods such as combining the HAM and Laplace
transformations (HATM) \cite{man6,HATM7,man1,man17}, optimal
homotopy analysis method \cite{ohm}, discrete homotopy analysis
method \cite{dhm1,dhm2}, predictor homotopy analysis method
\cite{phm1,phm2,phm3}, homotopy analysis Sumudu transform method
\cite{hasm1,hasm2} and many others \cite{nhm,man4,shm,chm}.

\begin{figure}
\centering
 \includegraphics[width=4in]{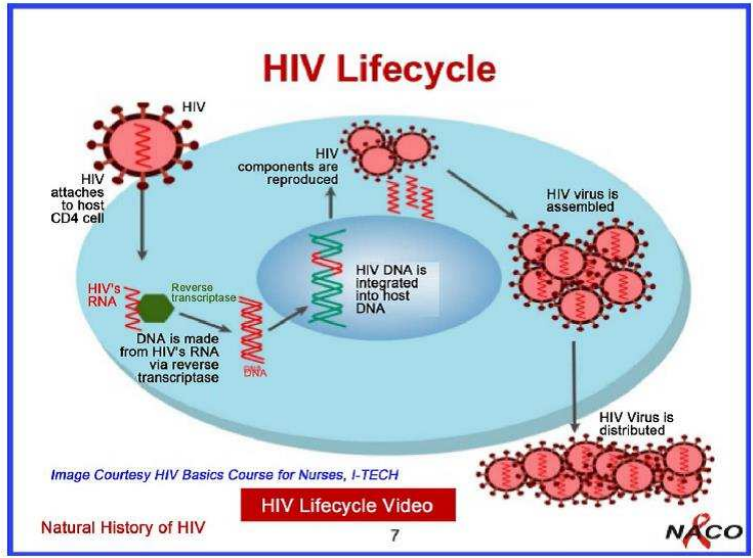}\\
  \caption{HIV life cycle.}\label{f00}
\end{figure}

\begin{figure}
\centering
  \includegraphics[width=4in]{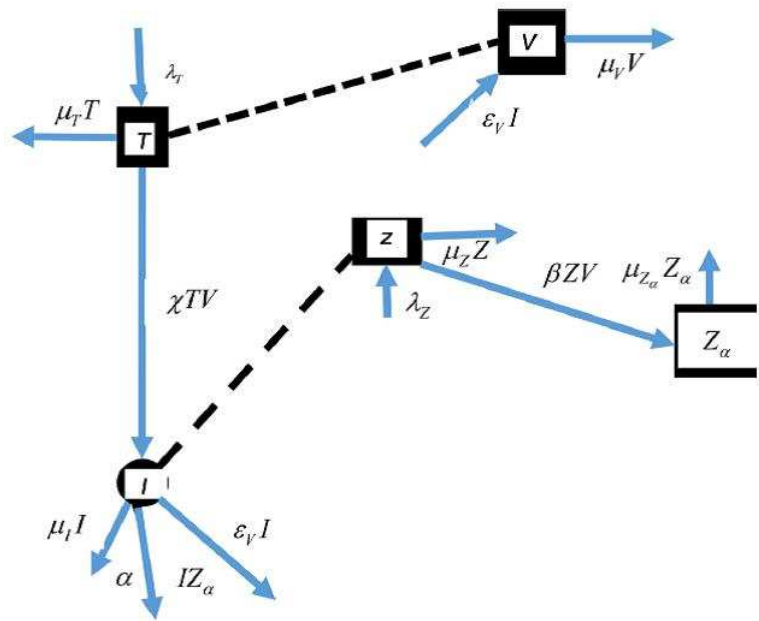}\\
  \caption{Diagram of HIV infection model of CD8$^+$ T-cells.}\label{f000}
\end{figure}

The aim of this paper is to present the HATM to solve the following
non-linear bio-mathematical model \cite{hiv10}
\begin{equation}\label{1}
\begin{array}{l}
\displaystyle \frac{dT(t)}{dt} = \lambda_T - \mu_T T(t) - \chi T(t)V(t)\\
\\
\displaystyle \frac{dI(t)}{dt} = \chi T(t)V(t) - \mu_I I(t) - \alpha I(t) Z_a(t),\\
\\
\displaystyle \frac{dV(t)}{dt} =\epsilon_V \mu_I I(t) - \mu_V V(t),\\
\\
\displaystyle \frac{dZ(t)}{dt} = \lambda_Z-\mu_Z Z(t) - \beta Z(t) I(t), \\
\\
\displaystyle \frac{dZ_a(t)}{dt} = \beta Z(t) I(t) - \mu_{Z_a}
Z_a(t),
\end{array}
\end{equation}
where $T(t)$ and $I(t)$ show the condensation of the susceptible and
infected CD4$^+$ T-cells at any time $t$,  $V(t)$ is the
condensation of infectious HIV viruses and finally $Z(t)$ and
$Z_a(t)$ are the condensation of the CD8$^+$ T-cells and population
of the activated CD8$^+$ T-cells at any time $t$. List of parameters
and their values are presented in Table \ref{t0}
\cite{hiv12,hiv13,hiv2,hiv3,hiv4,hiv11}. Moreover, in Figs.
\ref{f00} and \ref{f000} the life cycle of HIV infection and its
model on CD8$^+$ T-cells are demonstrated \cite{hiv10}.

The HATM obtains by combining the HAM with Laplace transformations.
Recently, the HATM has been applied to solve the various problems
such as solving singular problems \cite{man1}, fractional modeling
for BBM-Burger equation \cite{HATM1}, Klein-Gordon equations
\cite{HATM3}, fractional diffusion problem \cite{HATM4}, partial
differential equations \cite{HATM6}, fuzzy problems
\cite{man6,HATM5} and others \cite{hatm10,hatm11,HATM2}.

This research is organized in the following form: Section 2 is the
main idea for solving the non-linear bio-mathematical model \ref{1}.
The convergence theorem for solving presented model is illustrated
in Section 3. In Section 4, the numerical results for $N=5, 10$ are
presented. Also, several  $\hbar$-curves are demonstrated to show
the convergence regions of this problem. Furthermore, the plots of
residual error functions are presented to show the precision of
method. Finally, Section 5 is conclusion.

\begin{table}[h]
\caption{ List of parameters and their values. }\label{t0}
 \centering
\scalebox{0.7}{
\begin{tabular}{|c|l|c|}
\hline
 Parameters & ~~~~~~~~~~~~~~~~~~~~~~~~~~~~~~~~~~~~~~~~~~~~~~~~Meaning   & Values      \\
    \hline
$\lambda_T$&Rate of  recruiting  the susceptible CD4$^+$ T-cells per unit time.& 10 cell/mm$^3$/day\\
$\mu_T$&Rate of decaying for susceptible CD4$^+$ T-cells.&0.01 day$^{-1}$\\
$\chi$&Rate of infecting for CD4$^+$ T-cells by the virus.&0.000024 mm$^3$ vir$^{-1}$ day$^{-1}$\\
$\mu_I$&Rate of the natural death for infected CD4$^+$ T-cells.&0.5 day$^{-1}$\\
$\epsilon_V$& Rate of generation for HIV virions by infected CD4$^+$ T-cells.&100 vir. cell$^{-1}$ day$^{-1}$\\
$\mu_V$&Rate of the death for infectious virus.&3 day$^{-1}$\\
$\alpha$&Rate of eliminating the infected cells by the activated CD8$^+$ T-cells.&0.02 day$^{-1}$\\
$\lambda_Z$&Rate of  recruiting the CD8$^+$ T-cells per unit time.&20 cell/mm$^3$/day \\
$\mu_Z$&Rate of the death for CD8$^+$ T-cells.&0.06 day$^{-1}$\\
$\beta$&Rate of activation for CD8$^+$ T-cells due to the attendance the infected CD4$^+$ T-cells.&0.004 day$^{-1}$\\
$\mu_{Z_a}$&Rate of decaying for activated defence cells decay per unit time.&0.004 day$^{-1}$\\
 \hline
 \end{tabular}
 }
\end{table}

\section{Homotopy Analysis Transform Method}
Defining the linear operators $L_T, L_I, L_V, L_Z, L_{Z_a}$ as
follows
\begin{equation}\label{11}
\begin{array}{lllll}
\displaystyle   L_T =  L_I =  L_V =  L_Z =  L_{Z_a} = \mathcal{L}, \\
\end{array}
\end{equation}
where $\mathcal{L}$ is the Laplace transformation. Applying this
operator for non-linear system of Eqs. (\ref{1}) as
\begin{equation}\label{12}
\begin{array}{l}
\displaystyle \mathcal{L}[T(t)] =  \frac{T(0)}{s} + \frac{\mathcal{L}[\lambda_T]}{s} - \frac{\mu_T}{s}  \mathcal{L}[{T}(t)] - \frac{\chi}{s}  \mathcal{L}[{T}(t)  {V}(t)], \\
\\
\displaystyle \mathcal{L}[I(t)] =  \frac{I(0)}{s} + \frac{\chi}{s}
\mathcal{L}[{T}(t) {V}(t)] - \frac{\mu_I}{s}  \mathcal{L}[{I}(t)] - \frac{\alpha}{s}  \mathcal{L}[{I}(t)  {Z_a}(t)],  \\
\\
\displaystyle \mathcal{L}[V(t)] =  \frac{V(0)}{s} +\frac{\epsilon_V \mu_I}{s}  \mathcal{L}[{I}(t)] - \frac{\mu_V}{s}  \mathcal{L}[{V}(t)], \\
\\
\displaystyle \mathcal{L}[Z(t)] =  \frac{Z(0)}{s} + \frac{\mathcal{L}[\lambda_Z]}{s}+\frac{\mu_Z}{s}  \mathcal{L}[{Z}(t)] - \frac{\beta}{s}  \mathcal{L}[{Z}(t)  {I}(t)], \\
\\
\displaystyle \mathcal{L}[Z_a(t)] =   \frac{Z_a(0)}{s} +
\frac{\beta}{s}
 \mathcal{L}[{Z}(t)  {I}(t)] - \frac{\mu_{Z_a}}{s}
 \mathcal{L}[{Z_a}(t)]. \\
\\
\end{array}
\end{equation}

Let $0\leq q \leq 1$ be an embedding parameter, $\hbar$ is an
auxiliary parameter, $H_T(t), H_I(t), H_V(t), H_Z(t)$ and
$H_{Z_a}(t)$ are the auxiliary functions, $L_T, L_I, L_V, L_Z,
L_{Z_a}$ are the linear operators and $N_T, N_I, N_V, N_Z, N_{Z_a}$
are the non-linear operators then the following Homotopy maps can be
defined as
\begin{equation}\label{3}
\begin{array}{l}
H_T[\hat{T}(t;q), \hat{I}(t;q), \hat{V}(t;q), \hat{Z}(t;q),
\hat{Z_a}(t;q)] \\
\\
~~~= (1-q) L_T[\hat{T}(t;q)-T_0(t)] - q \hbar H_T(t)
N_T[\hat{T}(t;q), \hat{I}(t;q), \hat{V}(t;q), \hat{Z}(t;q),
\hat{Z_a}(t;q)],\\
\\
H_I[\hat{T}(t;q), \hat{I}(t;q), \hat{V}(t;q), \hat{Z}(t;q),
\hat{Z_a}(t;q)] \\
\\
~~~= (1-q) L_I[\hat{I}(t;q)-I_0(t)] - q \hbar H_I(t)
N_I[\hat{T}(t;q), \hat{I}(t;q), \hat{V}(t;q), \hat{Z}(t;q),
\hat{Z_a}(t;q)],\\
\\
H_V[\hat{T}(t;q), \hat{I}(t;q), \hat{V}(t;q), \hat{Z}(t;q),
\hat{Z_a}(t;q)] \\
\\
~~~= (1-q) L_V[\hat{V}(t;q)-V_0(t)] - q \hbar H_V(t)
N_V[\hat{T}(t;q), \hat{I}(t;q), \hat{V}(t;q), \hat{Z}(t;q),
\hat{Z_a}(t;q)],\\
\\
H_Z[\hat{T}(t;q), \hat{I}(t;q), \hat{V}(t;q), \hat{Z}(t;q),
\hat{Z_a}(t;q)] \\
\\
~~~= (1-q) L_Z[\hat{Z}(t;q)-Z_0(t)] - q \hbar H_Z(t)
N_Z[\hat{T}(t;q), \hat{I}(t;q), \hat{V}(t;q), \hat{Z}(t;q),
\hat{Z_a}(t;q)],\\
\\
H_{Z_a}[\hat{T}(t;q), \hat{I}(t;q), \hat{V}(t;q), \hat{Z}(t;q),
\hat{Z_a}(t;q)] \\
\\
~~~= (1-q) L_{Z_a}[\hat{Z_a}(t;q)-{Z_a}_0(t)] - q \hbar H_{Z_a}(t)
N_{Z_a}[\hat{T}(t;q), \hat{I}(t;q), \hat{V}(t;q), \hat{Z}(t;q),
\hat{Z_a}(t;q)],
\end{array}
\end{equation}
where the non-linear operators $N_T, N_I, N_V, N_Z, N_{Z_a}$ are
defined in the following forms
\begin{equation}\label{4}
\begin{array}{l}
\displaystyle N_T[\hat{T}(t;q), \hat{I}(t;q), \hat{V}(t;q),
\hat{Z}(t;q),
\hat{Z_a}(t;q)] =  \frac{\partial \hat{T}(t;q)}{\partial t} - \lambda_T + \mu_T \hat{T}(t;q) + \chi \hat{T}(t;q) \hat{V}(t;q), \\
\\
\displaystyle N_I[\hat{T}(t;q), \hat{I}(t;q), \hat{V}(t;q),
\hat{Z}(t;q), \hat{Z_a}(t;q)]=  \frac{\partial \hat{I}(t;q)}{\partial t} - \chi \hat{T}(t;q) \hat{V}(t;q) + \mu_I \hat{I}(t;q) + \alpha \hat{I}(t;q) \hat{Z_a}(t;q),  \\
\\
\displaystyle N_V[\hat{T}(t;q), \hat{I}(t;q), \hat{V}(t;q),
\hat{Z}(t;q),
\hat{Z_a}(t;q)]=  \frac{\partial \hat{V}(t;q)}{\partial t} -\epsilon_V \mu_I \hat{I}(t;q) + \mu_V \hat{V}(t;q), \\
\\
\displaystyle N_Z[\hat{T}(t;q), \hat{I}(t;q), \hat{V}(t;q),
\hat{Z}(t;q),
\hat{Z_a}(t;q)]=  \frac{\partial \hat{Z}(t;q)}{\partial t} - \lambda_Z+\mu_Z \hat{Z}(t;q) + \beta \hat{Z}(t;q) \hat{I}(t;q), \\
\\
\displaystyle N_{Z_a}[\hat{T}(t;q), \hat{I}(t;q), \hat{V}(t;q),
\hat{Z}(t;q), \hat{Z_a}(t;q)]=   \frac{\partial
\hat{Z_a}(t;q)}{\partial t} - \beta \hat{Z}(t;q) \hat{I}(t;q) +
\mu_{Z_a}
\hat{Z_a}(t;q). \\
\\
\end{array}
\end{equation}
Now, we can construct the following zero order deformation equations
as
\begin{equation}\label{5}
\begin{array}{l}
(1-q) L_T[\hat{T}(t;q)-T_0(t)] - q \hbar H_T(t) N_T[\hat{T}(t;q),
\hat{I}(t;q), \hat{V}(t;q), \hat{Z}(t;q),
\hat{Z_a}(t;q)]=0,\\
\\
(1-q) L_I[\hat{I}(t;q)-I_0(t)] - q \hbar H_I(t) N_I[\hat{T}(t;q),
\hat{I}(t;q), \hat{V}(t;q), \hat{Z}(t;q),
\hat{Z_a}(t;q)]=0,\\
\\
(1-q) L_V[\hat{V}(t;q)-V_0(t)] - q \hbar H_V(t) N_V[\hat{T}(t;q),
\hat{I}(t;q), \hat{V}(t;q), \hat{Z}(t;q),
\hat{Z_a}(t;q)]=0,\\
\\
(1-q) L_Z[\hat{Z}(t;q)-Z_0(t)] - q \hbar H_Z(t) N_Z[\hat{T}(t;q),
\hat{I}(t;q), \hat{V}(t;q), \hat{Z}(t;q),
\hat{Z_a}(t;q)]=0,\\
\\
(1-q) L_{Z_a}[\hat{Z_a}(t;q)-{Z_a}_0(t)] - q \hbar H_{Z_a}(t)
N_{Z_a}[\hat{T}(t;q), \hat{I}(t;q), \hat{V}(t;q), \hat{Z}(t;q),
\hat{Z_a}(t;q)]=0.
\end{array}
\end{equation}
Using the following Taylor expansions as
\begin{equation}\label{6}
\begin{array}{l}
\displaystyle \hat{T}(t;q) = T_0(t) + \sum_{m=1}^{\infty} T_m(t) q^m,\\
 \\
\displaystyle \hat{I}(t;q) = I_0(t) + \sum_{m=1}^{\infty} I_m(t) q^m,\\
\\
\displaystyle \hat{V}(t;q)= V_0(t) + \sum_{m=1}^{\infty} V_m(t) q^m,\\
 \\
\displaystyle  \hat{Z}(t;q)= Z_0(t) + \sum_{m=1}^{\infty} Z_m(t) q^m,\\
  \\
\displaystyle \hat{Z_a}(t;q)= {Z_a}_0(t) + \sum_{m=1}^{\infty}
{Z_a}_m(t) q^m,
\end{array}
\end{equation}
where
$$
\begin{array}{lll}
\displaystyle T_m=\frac{1}{m!} \frac{\partial^m
\hat{T}(t;q)}{\partial q^m}\bigg|_{q=0}, &\displaystyle
I_m=\frac{1}{m!} \frac{\partial^m \hat{I}(t;q)}{\partial
q^m}\bigg|_{q=0},&\displaystyle V_m=\frac{1}{m!} \frac{\partial^m
\hat{V}(t;q)}{\partial q^m}\bigg|_{q=0},\\
\\
\displaystyle Z_m=\frac{1}{m!} \frac{\partial^m
\hat{Z}(t;q)}{\partial q^m}\bigg|_{q=0},&\displaystyle
{Z_a}_m=\frac{1}{m!} \frac{\partial^m \hat{Z_a}(t;q)}{\partial
q^m}\bigg|_{q=0}.
\end{array}
$$

Defining the following vectors
$$
\begin{array}{l}
\displaystyle  \hat{T}_m(t) = \bigg \{ T_0(t), T_1(t), \ldots, T_m(t) \bigg\},  \\
\\
\displaystyle  \hat{I}_m(t) = \bigg \{ I_0(t), I_1(t), \ldots, I_m(t) \bigg\},  \\
  \\
\displaystyle  \hat{V}_m(t) = \bigg \{ V_0(t), V_1(t), \ldots, V_m(t) \bigg\},  \\
\\
\displaystyle  \hat{Z}_m(t) = \bigg \{ Z_0(t), Z_1(t), \ldots, Z_m(t) \bigg\},  \\
\\
\displaystyle  {\hat{Z_a}}_m(t) = \bigg \{ {Z_a}_0(t), {Z_a}_1(t), \ldots, {Z_a}_m(t) \bigg\},  \\
\end{array}
$$
to construct the $m$-th order deformation equations as follows
\begin{equation}\label{7}
    \begin{array}{l}
     \displaystyle L_T \left[T_m(t) - \chi_m T_{m-1}(t)\right] = \hbar H_T (t) \Re^T_{m} \left( \vec{T}_{m-1}, \vec{I}_{m-1}, \vec{V}_{m-1}, \vec{Z}_{m-1}, \vec{Z_a}_{m-1}   \right),\\
\\
     \displaystyle L_I \left[I_m(t) - \chi_m I_{m-1}(t)\right] = \hbar H_I (t) \Re^I_{m} \left( \vec{T}_{m-1}, \vec{I}_{m-1}, \vec{V}_{m-1}, \vec{Z}_{m-1}, \vec{Z_a}_{m-1}   \right),\\
\\
     \displaystyle L_V \left[V_m(t) - \chi_m V_{m-1}(t)\right] = \hbar H_V (t) \Re^V_{m} \left( \vec{T}_{m-1}, \vec{I}_{m-1}, \vec{V}_{m-1}, \vec{Z}_{m-1}, \vec{Z_a}_{m-1}   \right),\\
\\
     \displaystyle L_Z \left[Z_m(t) - \chi_m Z_{m-1}(t)\right] = \hbar H_Z (t) \Re^Z_{m} \left( \vec{T}_{m-1}, \vec{I}_{m-1}, \vec{V}_{m-1}, \vec{Z}_{m-1}, \vec{Z_a}_{m-1}   \right),\\
\\
     \displaystyle L_{Z_a} \left[{Z_a}_m(t) - \chi_m {Z_a}_{m-1}(t)\right] = \hbar H_{Z_a} (t) \Re^{Z_a}_{m} \left( \vec{T}_{m-1}, \vec{I}_{m-1}, \vec{V}_{m-1}, \vec{Z}_{m-1}, \vec{Z_a}_{m-1}   \right),\\
     \end{array}
\end{equation}
where
\begin{equation}\label{14}
\begin{array}{l}
\displaystyle \Re^T_{m} = \mathcal{L }[T_{m-1}] - \frac{T_{m-1(0)}}{s} - (1-\chi_m)\frac{\mathcal{L }[\lambda_T]}{s} + \frac{\mu_T}{s}  \mathcal{L }[{T_{m-1}}(t)] + \frac{\chi}{s} \mathcal{L }\left[\sum_{j=0}^{m-1} {T_j}(t)  {V_{m-1-j}}(t)\right],  \\
\\
\displaystyle \Re^I_{m}=   \mathcal{L }[I_{m-1}] - \frac{I_{m-1(0)}}{s} - \frac{\chi}{s} \mathcal{L }\left[\sum_{j=0}^{m-1} {T_j}(t)  {V_{m-1-j}}(t)\right] + \frac{\mu_I}{s}  \mathcal{L }[{I_{m-1}}(t)] + \frac{\alpha}{s}  \mathcal{L }\left[\sum_{j=0}^{m-1} {I_j}(t)  {Z_a}_{m-1-j}(t)\right],  \\
\\
\displaystyle \Re^V_{m}=  \mathcal{L }[V_{m-1}] - \frac{V_{m-1(0)}}{s} -\frac{\epsilon_V \mu_I }{s} \mathcal{L }[{I_{m-1}}(t)] + \frac{\mu_V }{s} \mathcal{L }[{V_{m-1}}(t)],  \\
\\
\displaystyle \Re^Z_{m}= \mathcal{L }[Z_{m-1}] - \frac{Z_{m-1(0)}}{s} -  (1-\chi_m)\frac{ \mathcal{L }[\lambda_Z]}{s}+\frac{\mu_Z}{s}  \mathcal{L }[{Z_{m-1}}(t)] + \frac{\beta}{s} \mathcal{L }\left[ \sum_{j=0}^{m-1} {Z_j}(t)  {I_{m-1-j}}(t)\right], \\
\\
\displaystyle \Re^{Z_a}_{m}=  \mathcal{L }[{Z_a}_{m-1}] -
\frac{{Z_a}_{m-1(0)}}{s} - \frac{\beta}{s} \mathcal{L }\left[
\sum_{j=0}^{m-1} {Z_j}(t) {I_{m-1-j}}(t)\right] +\frac{
\mu_{Z_a}}{s}
\mathcal{L }[ {Z_a}_{m-1}(t)], \\
\\
\end{array}
\end{equation}
and
\begin{equation}\label{khi}
\chi_m = \left\{ \begin{array}{l}
                   0, ~~~~~m\leq 1 \\
                     \\
                   1, ~~~~~m > 1.
                 \end{array}\right.
\end{equation}

Applying the inverse Laplace transformation $\mathcal{L}^{-1}$ for
Eqs. (\ref{7}) we get
\begin{equation}\label{15}
\begin{array}{c}
\displaystyle  T_m(t) =\chi_m T_{m-1} (t)+\hbar \mathcal{L}^{-1}\left[ \Re^{T}_{m} (t) \right],  \\
\\
\displaystyle  I_m(t) =\chi_m I_{m-1} (t)+\hbar \mathcal{L}^{-1}\left[ \Re^{I}_{m} (t)\right],  \\
\\
\displaystyle  V_m(t) =\chi_m V_{m-1} (t)+\hbar \mathcal{L}^{-1}\left[ \Re^{V}_{m} (t) \right],  \\
\\
\displaystyle  Z_m(t) =\chi_m Z_{m-1} (t)+\hbar \mathcal{L}^{-1}\left[ \Re^{Z}_{m} (t) \right],  \\
\\
\displaystyle  {Z_a}_m(t) =\chi_m {Z_a}_{m-1} (t)+\hbar \mathcal{L}^{-1}\left[ \Re^{Z_a}_{m} (t) \right],  \\
\end{array}
\end{equation}
and finally the approximate solutions can be obtained by
\begin{equation}\label{10}
\begin{array}{lll}
\displaystyle T_m(t) = \sum_{j=0}^{m} T_j(t),~~~~~ & \displaystyle
I_m(t) = \sum_{j=0}^{m} I_j(t),~~~~~ & \displaystyle V_m(t) =
\sum_{j=0}^{m} V_j(t),\\
\\
\displaystyle Z_m(t) = \sum_{j=0}^{m} Z_j(t),~~~~~ & \displaystyle
{Z_a}_m(t) = \sum_{j=0}^{m} {Z_a}_j(t).
\end{array}
\end{equation}

\section{Convergence Theorem}
By proving the following theorem, we can show the capabilities of
the HATM to solve the non-linear system of Eqs. (\ref{1}).

\begin{theorem} Let series solutions (\ref{10}) be convergent that are constructed by the $m$-th order deformation Eqs.
(\ref{7}). They must be the exact solution of system (\ref{1}).
\end{theorem}

\textbf{Proof:} Let the series solutions (\ref{10}) be convergent.
Hence, if
\begin{equation}\label{16}
\begin{array}{lll}
\displaystyle P_1(t) = \sum_{m=0}^{\infty} T_m(t),& \displaystyle
P_2(t) = \sum_{m=0}^{\infty} I_m(t),&
\displaystyle P_3(t) = \sum_{m=0}^{\infty} V_m(t),\\
\\
\displaystyle P_4(t) = \sum_{m=0}^{\infty} Z_m(t),& \displaystyle
P_5(t) = \sum_{m=0}^{\infty} {Z_a}_m(t),
\end{array}
\end{equation}
then
\begin{equation}\label{17}
\begin{array}{lll}
\lim_{m \rightarrow \infty} T_m(t) = 0,\\
\\
\lim_{m \rightarrow \infty} I_m(t) = 0,\\
\\
\lim_{m \rightarrow \infty} V_m(t) = 0,\\
\\
\lim_{m \rightarrow \infty} Z_m(t) = 0,\\
\\
\lim_{m \rightarrow \infty} {Z_a}_m(t) = 0.\\
\end{array}
\end{equation}
So, we can write
\begin{equation}\label{18}
\begin{array}{l}
\displaystyle  \sum_{m=1}^N \bigg[T_m(t) - \chi_m T_{m-1}(t)\bigg] =  T_N(t),\\
  \\
\displaystyle  \sum_{m=1}^N \bigg[I_m(t) - \chi_m I_{m-1}(t)\bigg] = I_N(t),\\
  \\
\displaystyle  \sum_{m=1}^N \bigg[V_m(t) - \chi_m v_{m-1}(t)\bigg]
=V_N(t),\\
\\
\displaystyle  \sum_{m=1}^N \bigg[Z_m(t) - \chi_m Z_{m-1}(t)\bigg]
=Z_N(t),\\
\\
\displaystyle  \sum_{m=1}^N \bigg[{Z_a}_m(t) - \chi_m
{Z_a}_{m-1}(t)\bigg]
={Z_a}_N(t),\\
\\
\end{array}
\end{equation}
where Eqs. (\ref{17}) and (\ref{18}) are applied to construct the
following relations as follows
\begin{equation}\label{19}
\begin{array}{l}
\displaystyle  \sum_{m=1}^N \bigg[T_m(t) - \chi_m T_{m-1}(t)\bigg] = \lim_{N \rightarrow \infty} T_N(t) = 0,\\
  \\
\displaystyle  \sum_{m=1}^N \bigg[I_m(t) - \chi_m I_{m-1}(t)\bigg] =
\lim_{N \rightarrow \infty}
I_N(t) = 0,\\
  \\
\displaystyle  \sum_{m=1}^N \bigg[V_m(t) - \chi_m V_{m-1}(t)\bigg] =
\lim_{N \rightarrow \infty} V_N(t) = 0,\\
  \\
\displaystyle  \sum_{m=1}^N \bigg[Z_m(t) - \chi_m Z_{m-1}(t)\bigg] =
\lim_{N \rightarrow \infty} Z_N(t) = 0,\\
  \\
\displaystyle  \sum_{m=1}^N \bigg[{Z_a}_m(t) - \chi_m
{Z_a}_{m-1}(t)\bigg] =
\lim_{N \rightarrow \infty} {Z_a}_N(t) = 0.\\
\end{array}
\end{equation}
Applying the linear operators $L_T, L_I, L_V, L_Z$ and $L_{Z_a}$ as
\begin{equation}\label{20}
\begin{array}{l}
\displaystyle  \sum_{m=1}^\infty L_T \bigg[T_m(t) - \chi_m T_{m-1}(t)\bigg] = L_T \bigg[ \sum_{m=1}^\infty T_m(t) - \chi_m T_{m-1}(t)\bigg]= 0 ,\\
  \\
\displaystyle  \sum_{m=1}^\infty L_I \bigg[I_m(t) - \chi_m I_{m-1}(t)\bigg] =  L_I \bigg[ \sum_{m=1}^\infty I_m(t) - \chi_m I_{m-1}(t)\bigg]= 0,\\
  \\
\displaystyle  \sum_{m=1}^\infty L_V \bigg[V_m(t) - \chi_m
V_{m-1}(t)\bigg] =  L_V \bigg[ \sum_{m=1}^\infty V_m(t) - \chi_m
V_{m-1}(t)\bigg]= 0,\\
  \\
\displaystyle  \sum_{m=1}^\infty L_Z \bigg[Z_m(t) - \chi_m
Z_{m-1}(t)\bigg] =  L_Z \bigg[ \sum_{m=1}^\infty Z_m(t) - \chi_m
Z_{m-1}(t)\bigg]= 0,\\
  \\
\displaystyle  \sum_{m=1}^\infty L_{Z_a} \bigg[{Z_a}_m(t) - \chi_m
{Z_a}_{m-1}(t)\bigg] =  L_{Z_a} \bigg[ \sum_{m=1}^\infty {Z_a}_m(t)
- \chi_m {Z_a}_{m-1}(t)\bigg]= 0.
\end{array}
\end{equation}
By using Eqs. (\ref{7}) and (\ref{20}) we get
\begin{equation}\label{21}
\begin{array}{l}
\displaystyle   \hbar H_T(t)  \sum_{m=1}^\infty \Re^T_{m} (\vec{T}_{m-1}, \vec{I}_{m-1}, \vec{V}_{m-1}, \vec{Z}_{m-1}, \vec{Z_a}_{m-1} ) =0,\\
  \\
\displaystyle   \hbar H_I(t)  \sum_{m=1}^\infty \Re^I_{m} (\vec{T}_{m-1}, \vec{I}_{m-1}, \vec{V}_{m-1}, \vec{Z}_{m-1}, \vec{Z_a}_{m-1} ) =0,\\
  \\
\displaystyle   \hbar H_V(t)  \sum_{m=1}^\infty \Re^V_{m} (\vec{T}_{m-1}, \vec{I}_{m-1}, \vec{V}_{m-1}, \vec{Z}_{m-1}, \vec{Z_a}_{m-1} ) =0,\\
  \\
\displaystyle   \hbar H_Z(t)  \sum_{m=1}^\infty \Re^Z_{m} (\vec{T}_{m-1}, \vec{I}_{m-1}, \vec{V}_{m-1}, \vec{Z}_{m-1}, \vec{Z_a}_{m-1} ) =0,\\
  \\
\displaystyle   \hbar H_{Z_a}(t)  \sum_{m=1}^\infty \Re^{Z_a}_{m} (\vec{T}_{m-1}, \vec{I}_{m-1}, \vec{V}_{m-1}, \vec{Z}_{m-1}, \vec{Z_a}_{m-1} ) =0.\\
\end{array}
\end{equation}

According to the base definitions of the HAM in Eqs. (\ref{21}),
$\hbar \neq 0, H_S(t)\neq 0, H_I(t)\neq 0, H_R(t)\neq 0$, thus
\begin{equation}\label{22}
\begin{array}{l}
\displaystyle     \sum_{m=1}^\infty \Re^T_{m} (\vec{T}_{m-1}, \vec{I}_{m-1}, \vec{V}_{m-1}, \vec{Z}_{m-1}, \vec{Z_a}_{m-1} ) =0,\\
  \\
\displaystyle    \sum_{m=1}^\infty \Re^I_{m} (\vec{T}_{m-1}, \vec{I}_{m-1}, \vec{V}_{m-1}, \vec{Z}_{m-1}, \vec{Z_a}_{m-1} ) =0,\\
  \\
\displaystyle    \sum_{m=1}^\infty \Re^V_{m} (\vec{T}_{m-1}, \vec{I}_{m-1}, \vec{V}_{m-1}, \vec{Z}_{m-1}, \vec{Z_a}_{m-1} ) =0,\\
  \\
\displaystyle    \sum_{m=1}^\infty \Re^Z_{m} (\vec{T}_{m-1}, \vec{I}_{m-1}, \vec{V}_{m-1}, \vec{Z}_{m-1}, \vec{Z_a}_{m-1} ) =0,\\
  \\
\displaystyle    \sum_{m=1}^\infty \Re^{Z_a}_{m} (\vec{T}_{m-1}, \vec{I}_{m-1}, \vec{V}_{m-1}, \vec{Z}_{m-1}, \vec{Z_a}_{m-1} ) =0.\\
\end{array}
\end{equation}
Substituting $\Re^T_{m}, \Re^I_{m}, \Re^V_{m}, \Re^Z_{m}$ and
$\Re^{Z_a}_{m}$ into Eqs. (\ref{22}) and assuming
$(.)'=\frac{d}{dt}$ the following formulas are obtained as
\begin{equation}\label{23}
\begin{array}{l}
\displaystyle \sum_{m=1}^{\infty} \Re^T_{m} = \sum_{m=1}^{\infty} \left[T'_{m-1} - (1-\chi_m)\lambda_T + \mu_T  {T_{m-1}}(t) + \chi \sum_{j=0}^{m-1} {T_j}(t)  {V_{m-1-j}}(t)\right]  \\
\\
\displaystyle = \sum_{m=0}^{\infty} T'_{m} -\lambda_T + \mu_T  \sum_{m=0}^{\infty} {T_{m}}(t) + \chi \sum_{m=1}^{\infty} \sum_{j=0}^{m-1} {T_j}(t)  {V_{m-1-j}}(t)  \\
\\
\displaystyle = \sum_{m=0}^{\infty} T'_{m} -\lambda_T + \mu_T  \sum_{m=0}^{\infty} {T_{m}}(t) + \chi  \sum_{j=0}^{\infty} \sum_{m=j+1}^{\infty} {T_j}(t)  {V_{m-1-j}}(t)  \\
\\
\displaystyle = \sum_{m=0}^{\infty} T'_{m} -\lambda_T + \mu_T  \sum_{m=0}^{\infty} {T_{m}}(t) + \chi  \sum_{j=0}^{\infty} {T_j}(t) \sum_{m=0}^{\infty}   {V_{m}}(t) \\
\\
\displaystyle = P'_1(t) -\lambda_T + \mu_T P_1(t) + \chi P_1(t)
P_3(t),
\end{array}
\end{equation}
and
\begin{equation}\label{24}
\begin{array}{l}
\displaystyle \sum_{m=1}^{\infty} \Re^I_{m}= \sum_{m=1}^{\infty}  \left[ I'_{m-1}  - \chi \sum_{j=0}^{m-1} {T_j}(t)  {V_{m-1-j}}(t) + \mu_I  {I_{m-1}}(t) + \alpha  \sum_{j=0}^{m-1} {I_j}(t)  {Z_a}_{m-1-j}(t)\right]  \\
\\
\displaystyle =   \sum_{m=0}^{\infty}  I'_{m}  - \chi \sum_{m=1}^{\infty}  \sum_{j=0}^{m-1} {T_j}(t)  {V_{m-1-j}}(t) + \mu_I \sum_{m=0}^{\infty}  {I_{m}}(t) + \alpha  \sum_{m=1}^{\infty}  \sum_{j=0}^{m-1} {I_j}(t)  {Z_a}_{m-1-j}(t)   \\
\\
\displaystyle =   \sum_{m=0}^{\infty}  I'_{m}  - \chi  \sum_{j=0}^{\infty} \sum_{m=j+1}^{\infty}  {T_j}(t)  {V_{m-1-j}}(t) + \mu_I \sum_{m=0}^{\infty}  {I_{m}}(t) + \alpha  \sum_{j=0}^{\infty} \sum_{m=j+1}^{\infty} {I_j}(t)  {Z_a}_{m-1-j}(t)   \\
\\
\displaystyle =   \sum_{m=0}^{\infty}  I'_{m}  - \chi  \sum_{j=0}^{\infty}  {T_j}(t) \sum_{m=0}^{\infty}   {V_{m}}(t) + \mu_I \sum_{m=0}^{\infty}  {I_{m}}(t) + \alpha  \sum_{j=0}^{\infty}  {I_j}(t) \sum_{m=0}^{\infty}  {Z_a}_{m}(t)   \\
\\
\displaystyle =   P'_2(t)  - \chi  P_1(t) P_3(t) + \mu_I P_2(t) + \alpha  P_2(t) P_5(t),   \\
\end{array}
\end{equation}
and
\begin{equation}\label{25}
\begin{array}{l}
\displaystyle \sum_{m=1}^{\infty} \Re^V_{m}=\sum_{m=1}^{\infty} \left[ V'_{m-1}  -\epsilon_V \mu_I  {I_{m-1}}(t) + \mu_V  {V_{m-1}}(t)\right]~~~~~~~~~~ \\
\\
\displaystyle =\sum_{m=0}^{\infty}  V'_{m}  -\epsilon_V \mu_I  \sum_{m=0}^{\infty} {I_{m}}(t) + \mu_V  \sum_{m=0}^{\infty} {V_{m}}(t) \\
\\
\displaystyle = P'_3(t)  -\epsilon_V \mu_I P_2(t) + \mu_V P_3(t),
\end{array}
\end{equation}
and
\begin{equation}\label{26}
\begin{array}{l}
\displaystyle \sum_{m=1}^{\infty} \Re^Z_{m}= \sum_{m=1}^{\infty} \left[ Z'_{m-1}  -  (1-\chi_m) \lambda_Z+\mu_Z  {Z_{m-1}}(t) + \beta \sum_{j=0}^{m-1} {Z_j}(t)  {I_{m-1-j}}(t)\right] \\
\\
\displaystyle = \sum_{m=0}^{\infty}  Z'_{m}  -  \lambda_Z+\mu_Z  \sum_{m=0}^{\infty} {Z_{m}}(t) + \beta \sum_{j=0}^{\infty} \sum_{m=j+1}^{\infty}  {Z_j}(t)  {I_{m-1-j}}(t) \\
\\
\displaystyle = \sum_{m=0}^{\infty}  Z'_{m}  -  \lambda_Z+\mu_Z  \sum_{m=0}^{\infty} {Z_{m}}(t) + \beta \sum_{j=0}^{\infty} {Z_j}(t) \sum_{m=0}^{\infty}    {I_{m}}(t) \\
\\
\displaystyle = P'_4(t) -  \lambda_Z+\mu_Z P_4(t) + \beta   P_4(t)
P_2(t),
\end{array}
\end{equation}
and
\begin{equation}\label{27}
\begin{array}{l}
\displaystyle \sum_{m=1}^{\infty} \Re^{Z_a}_{m}= \sum_{m=1}^{\infty}
\left[ {Z_a}'_{m-1}  - \beta \sum_{j=0}^{m-1} {Z_j}(t)
{I_{m-1-j}}(t) + \mu_{Z_a}
 {Z_a}_{m-1}(t) \right] \\
 \\
\displaystyle = \sum_{m=0}^{\infty}
 {Z_a}'_{m}  - \beta \sum_{m=1}^{\infty} \sum_{j=0}^{m-1} {Z_j}(t)
{I_{m-1-j}}(t) + \mu_{Z_a} \sum_{m=0}^{\infty}
 {Z_a}_{m}(t)\\
 \\
\displaystyle  = \sum_{m=0}^{\infty}
 {Z_a}'_{m}  - \beta \sum_{j=0}^{\infty} \sum_{m=j+1}^{\infty}  {Z_j}(t)
{I_{m-1-j}}(t) + \mu_{Z_a} \sum_{m=0}^{\infty}
 {Z_a}_{m}(t)\\
 \\
\displaystyle  = \sum_{m=0}^{\infty}
 {Z_a}'_{m}  - \beta \sum_{j=0}^{\infty} {Z_j}(t) \sum_{m=0}^{\infty}
{I_{m}}(t) + \mu_{Z_a} \sum_{m=0}^{\infty}
 {Z_a}_{m}(t)\\
 \\
\displaystyle = P'_5(t) - \beta P_4(t) P_2(t) + \mu_{Z_a} P_5(t).
\end{array}
\end{equation}

Eqs. (\ref{23}), (\ref{24}), (\ref{25}), (\ref{26}) and (\ref{27})
show that the series solutions $P_1(t), P_2(t), P_3(t), P_4(t)$ and
$P_5(t)$ must be the exact solutions of Eqs. (\ref{1}).

\section{Numerical Illustration}
In this section, in order to show the flexibility of HATM to solve
the non-linear bio-mathematical model (\ref{1}), the numerical
solutions for $N=5$ are presented as follows
$$
\begin{array}{ll}
T_5(t) =& 1000 + 0.12 \hbar t + 0.24 \hbar^2 t + 0.24 \hbar^3 t +
0.12 \hbar^4 t +
 0.024 \hbar^5 t + 0.361203 \hbar^2 t^2 \\
 \\
 & + 0.722406 \hbar^3 t^2 +
 0.541804 \hbar^4 t^2 + 0.144481 \hbar^5 t^2 + 0.409213 \hbar^3 t^3 +
 0.613819 \hbar^4 t^3  \\
 \\
 & + 0.245528 \hbar^5 t^3 + 0.17452 \hbar^4 t^4 +
 0.139616 \hbar^5 t^4 + 0.0238202 \hbar^5 t^5, \\
 \\
I_5(t) =& -0.12 \hbar t - 0.24 \hbar^2 t - 0.24 \hbar^3 t - 0.12
\hbar^4 t - 0.024
\hbar^5 t - 0.420003 \hbar^2 t^2 \\
\\
&- 0.840006 \hbar^3 t^2 - 0.630004 \hbar^4 t^2 -
 0.168001 \hbar^5 t^2 - 0.478009 \hbar^3 t^3 - 0.717014 \hbar^4 t^3  \\
 \\
 &-
 0.286805 \hbar^5 t^3 - 0.203898 \hbar^4 t^4 - 0.163119 \hbar^5 t^4 -
 0.0278351 \hbar^5 t^5,\\
 \\
 V_5(t) =& 1 + 15 \hbar t + 30. \hbar^2 t + 30. \hbar^3 t + 15. \hbar^4 t + 3. \hbar^5 t +
 51. \hbar^2 t^2+ 102. \hbar^3 t^2  \\
 \\
 &+ 76.5 \hbar^4 t^2 + 20.4 \hbar^5 t^2 +
 58. \hbar^3 t^3 + 87.0001 \hbar^4 t^3 + 34.8 \hbar^5 t^3 \\
 \\
 & + 24.7376 \hbar^4 t^4 +
 19.7901 \hbar^5 t^4 + 3.37631 \hbar^5 t^5,\\
 \\
 Z_5(t) =& 500 + 50. \hbar t + 100. \hbar^2 t + 100. \hbar^3 t + 50. \hbar^4 t + 10. \hbar^5 t +
 2.76 \hbar^2 t^2 + 5.52 \hbar^3 t^2 \\
 \\
 & + 4.14 \hbar^4 t^2 + 1.104 \hbar^5 t^2 -
 0.228002 \hbar^3 t^3 - 0.342003 \hbar^4 t^3 - 0.136801 \hbar^5 t^3 \\
 \\
 & -
 0.123345 \hbar^4 t^4 - 0.0986763 \hbar^5 t^4 - 0.0169991 \hbar^5 t^5,\\
 \\
{Z_a}_5(t) =& 0.24 \hbar^2 t^2 + 0.48 \hbar^3 t^2 + 0.36 \hbar^4 t^2
+ 0.096 \hbar^5
t^2 + 0.283522 \hbar^3 t^3 + 0.425283 \hbar^4 t^3 \\
\\
& + 0.170113 \hbar^5 t^3 +
 0.121777 \hbar^4 t^4 + 0.0974217 \hbar^5 t^4 + 0.0167226 \hbar^5 t^5,
 \end{array}
$$

\begin{figure}
\centering
$$\begin{array}{ccc}
 \includegraphics[width=2.2in]{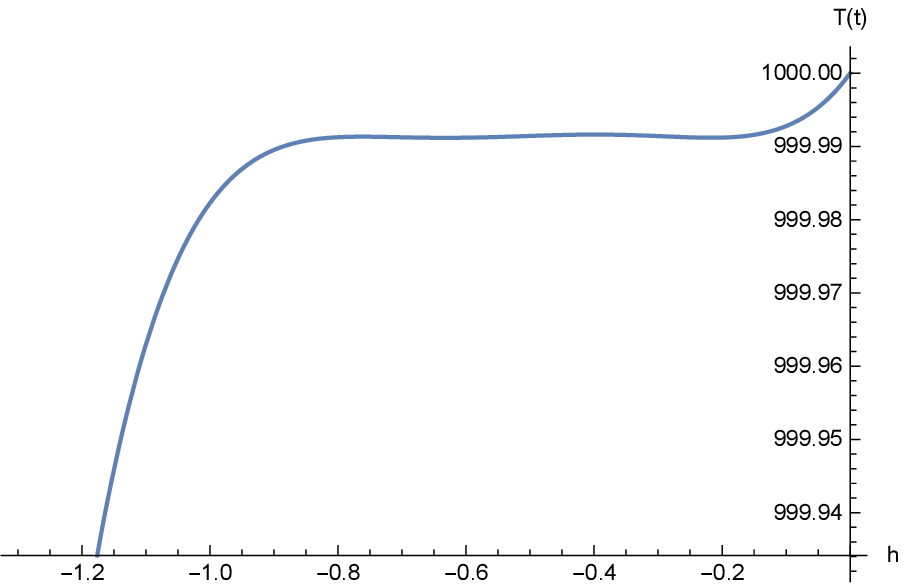}&~~~~~~~~&
 \includegraphics[width=2in]{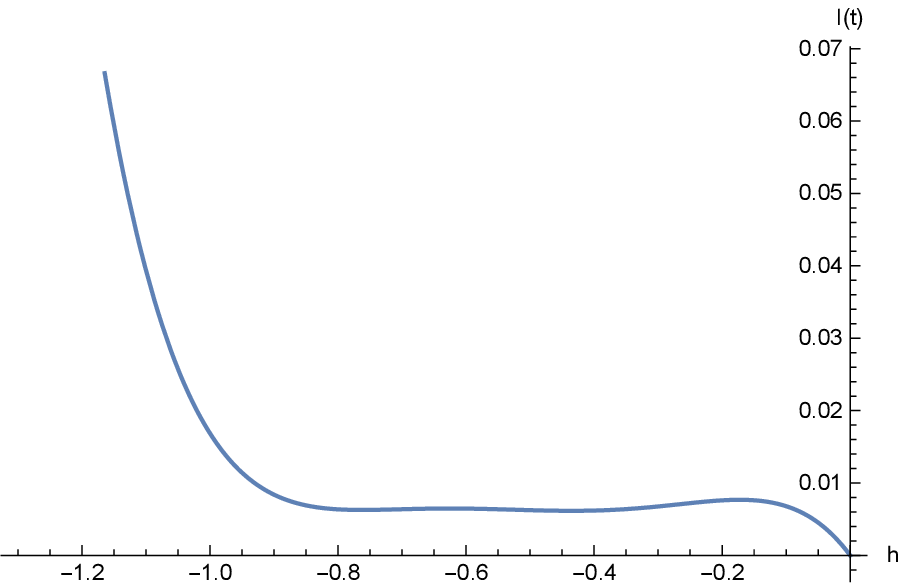}\\
 \\
 \\
 \includegraphics[width=2.2in]{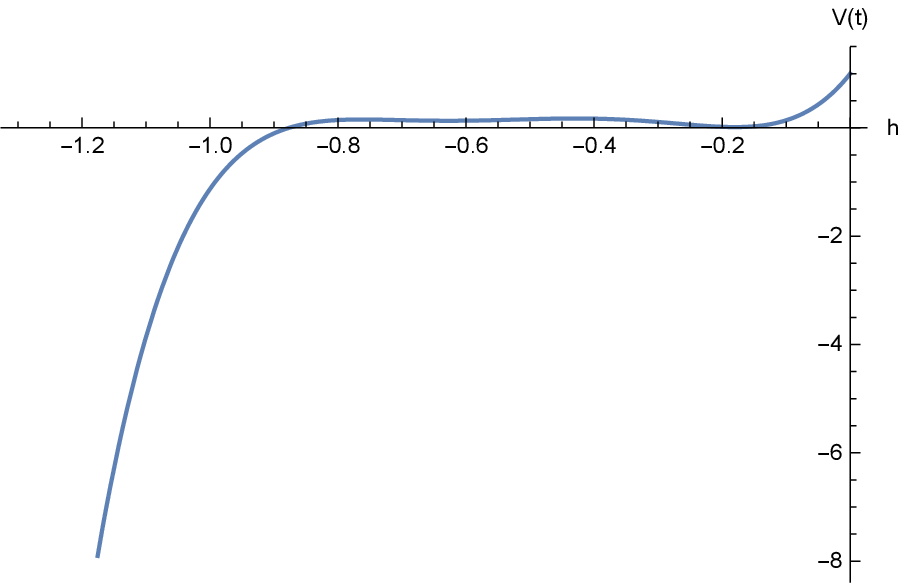}&~~~~~~~~&
 \includegraphics[width=2.2in]{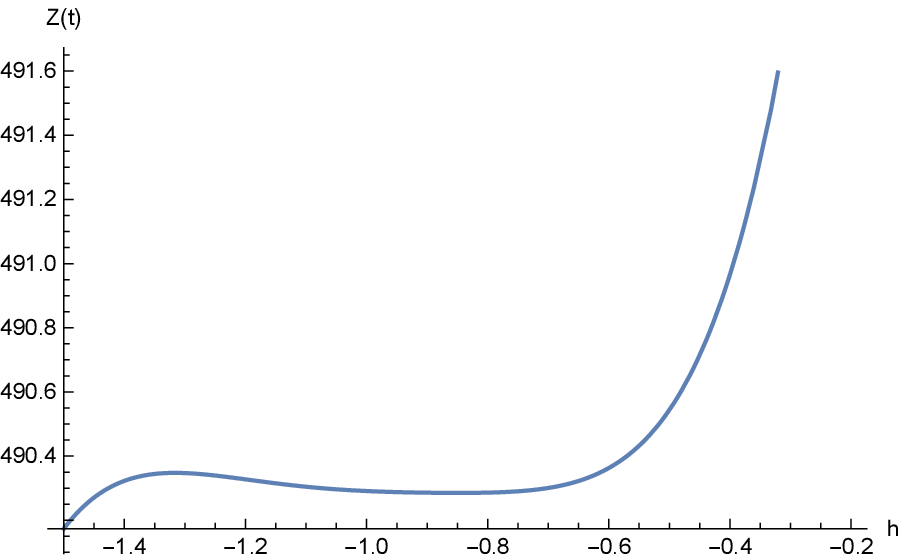}\\
  \\
  \\
 \includegraphics[width=2.2in]{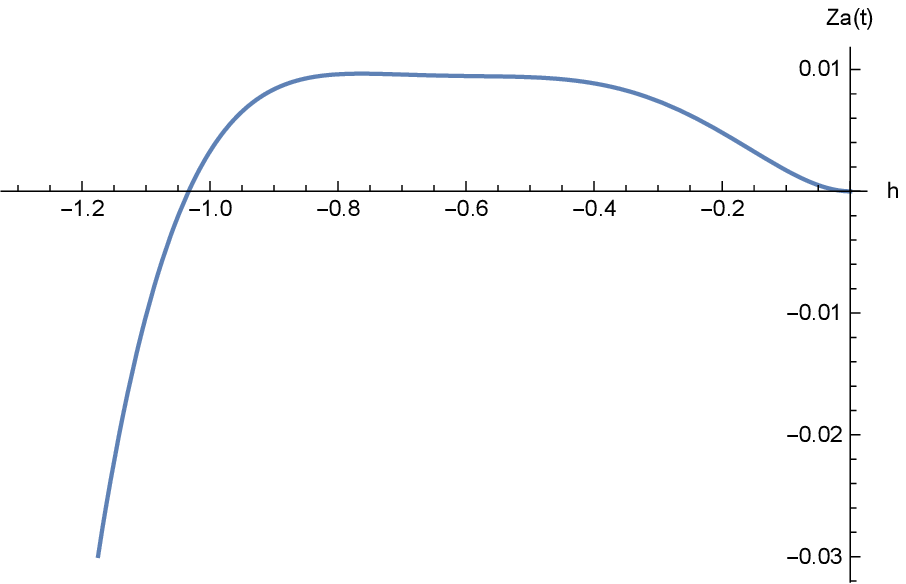}\\
\end{array}$$
  \caption{$\hbar$-curves of $T(t), I(t), V(t), Z(t)$ and $Z_a(t)$ for $N=5, t=1$. }\label{f1}
\end{figure}

\begin{figure}
\centering
$$\begin{array}{ccc}
 \includegraphics[width=2.2in]{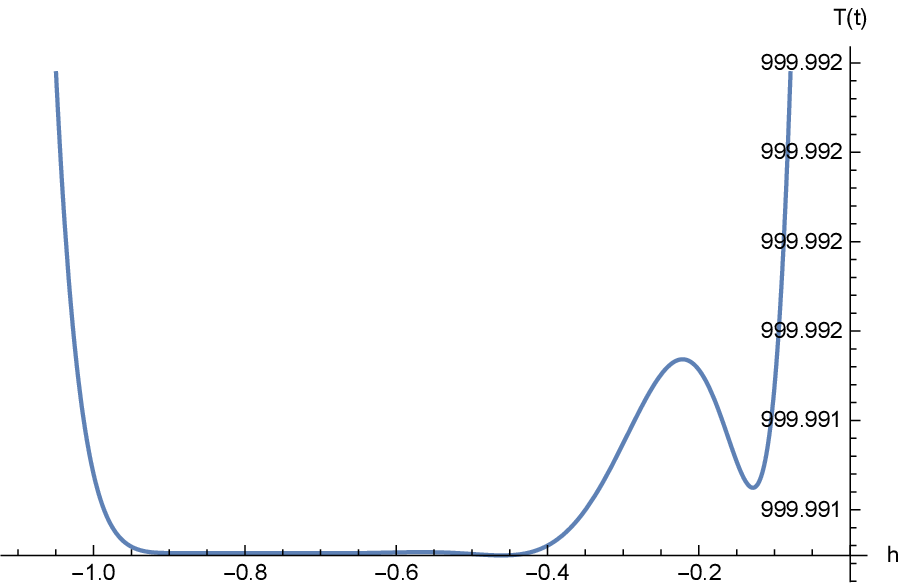}&~~~~~~~~&
 \includegraphics[width=2in]{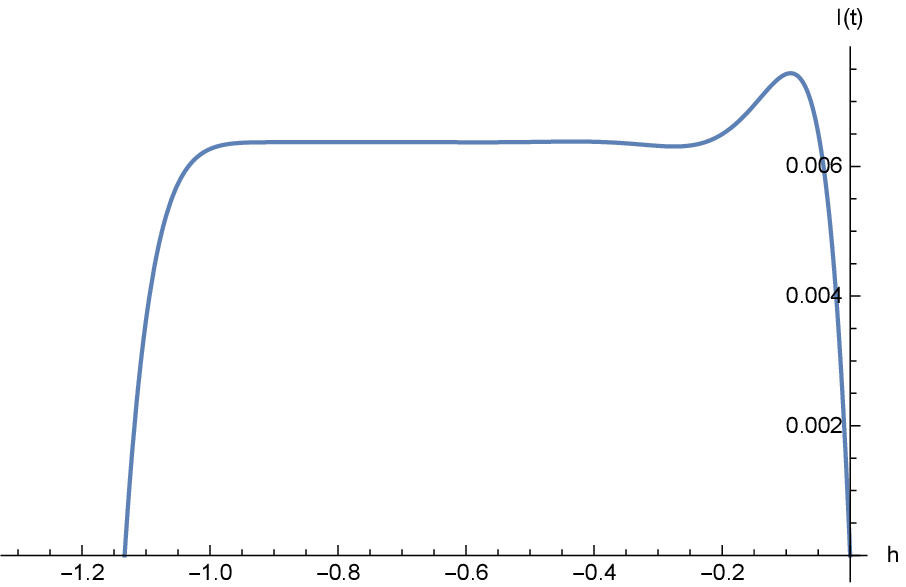}\\
 \\
 \\
 \includegraphics[width=2.2in]{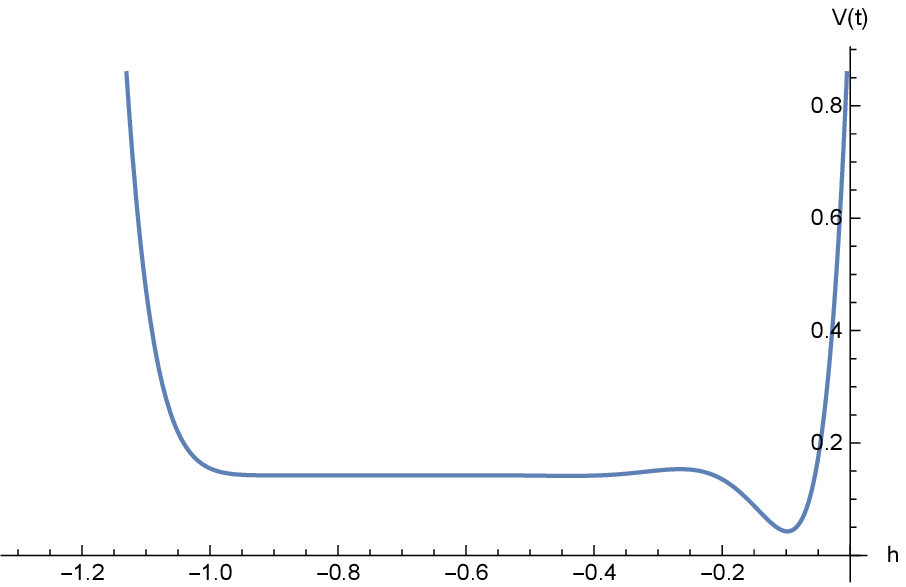}&~~~~~~~~&
 \includegraphics[width=2.2in]{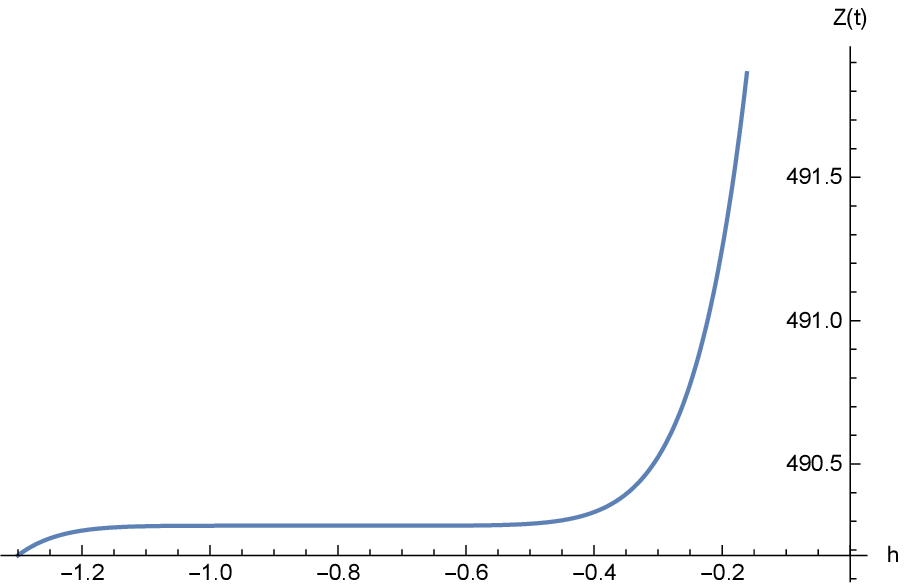}\\
  \\
  \\
 \includegraphics[width=2.2in]{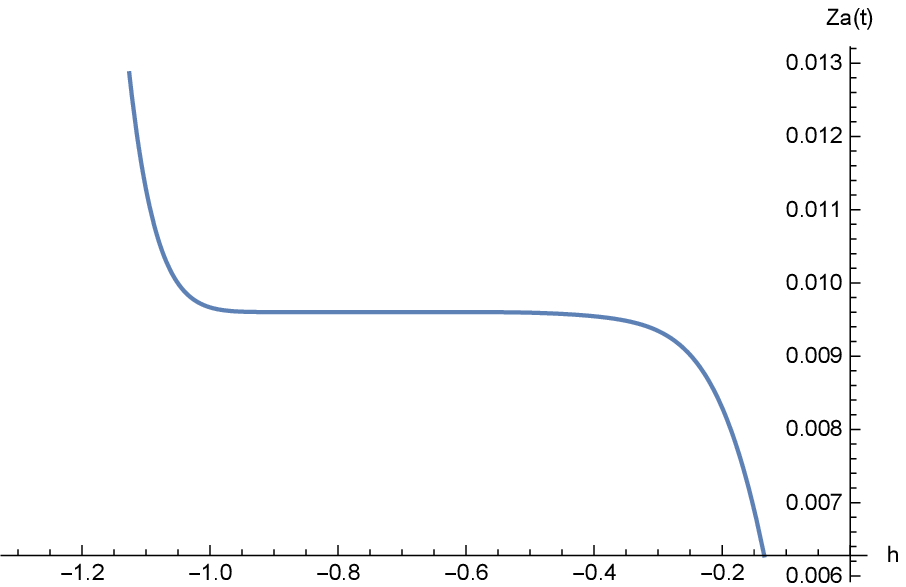}\\
\end{array}$$
  \caption{$\hbar$-curves of $T(t), I(t), V(t), Z(t)$ and $Z_a(t)$ for $N=10, t=1$. }\label{f2}
\end{figure}

The regions of convergence are shown by several $\hbar$-curves for
$N=5,10$ and $t=1$ in Figs. \ref{f1} and \ref{f2}. These regions are
parallel parts of $\hbar$-curves with axiom $x$. So for $N=5$ and
$t=1$ the convergence regions are
$$
\begin{array}{c}
 -0.9 \leq \hbar_T \leq -0.2, \\
 -0.8 \leq \hbar_I \leq -0.2, \\
 -0.8 \leq \hbar_V \leq -0.2, \\
 -1.2 \leq \hbar_Z \leq -0.6, \\
 -0.8 \leq \hbar_{Z_a} \leq -0.4, \\
\end{array}
$$
and for $N=10$ we get
$$
\begin{array}{c}
 -0.9 \leq \hbar_T \leq -0.4, \\
 -1 \leq \hbar_I \leq -0.2, \\
 -1 \leq \hbar_V \leq -0.3, \\
 -1.2 \leq \hbar_Z \leq -0.4, \\
 -1 \leq \hbar_{Z_a} \leq -0.4. \\
\end{array}
$$

\begin{figure}
\centering
$$\begin{array}{c}
 \includegraphics[width=3in]{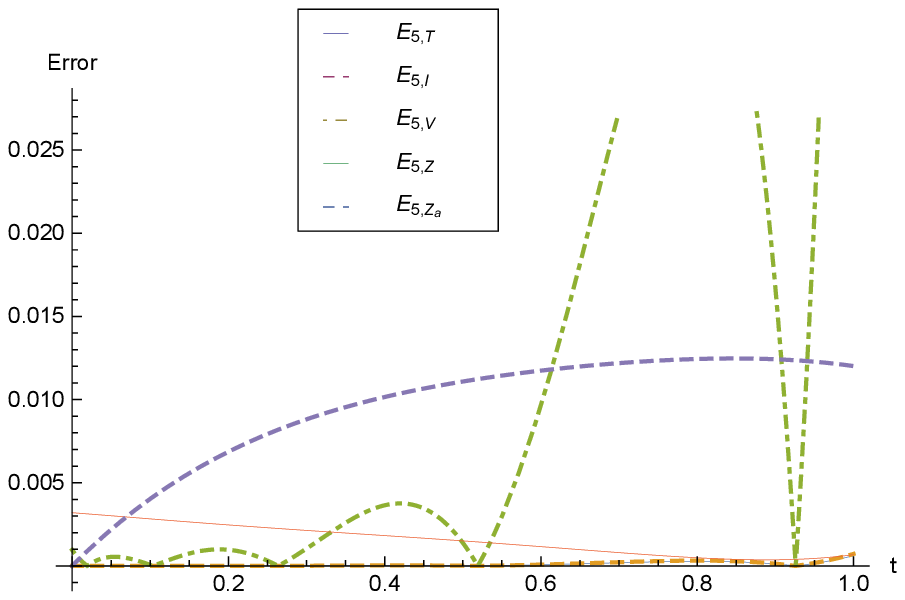}\\
 \\
 \\
 \includegraphics[width=3in]{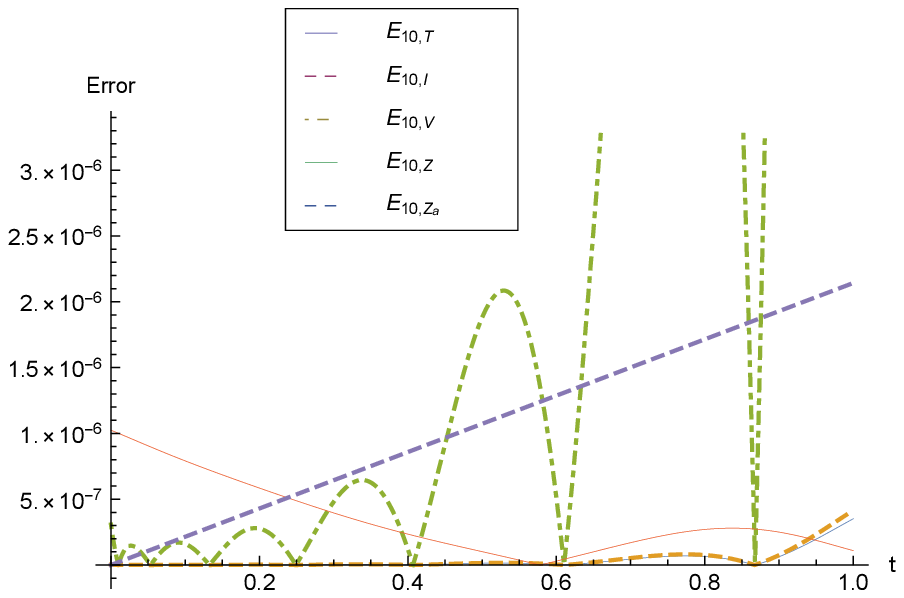}\\
\end{array}$$
  \caption{Residual error functions for $N=5,10$ and $\hbar=0.8$. }\label{f3}
\end{figure}

Also, the following residual error functions are applied to show the
accuracy of presented method as
\begin{equation}\label{28}
\begin{array}{l}
\displaystyle E_{N,T}(t) = \frac{dT(t)}{dt} - \lambda_T + \mu_T T(t) + \chi T(t)V(t)\\
\\
\displaystyle E_{N,I}(t) =  \frac{dI(t)}{dt} - \chi T(t)V(t) + \mu_I I(t) + \alpha I(t) Z_a(t),\\
\\
\displaystyle E_{N,V}(t) = \frac{dV(t)}{dt} -\epsilon_V \mu_I I(t) + \mu_V V(t),\\
\\
\displaystyle E_{N,Z}(t) = \frac{dZ(t)}{dt} - \lambda_Z+\mu_Z Z(t) + \beta Z(t) I(t), \\
\\
\displaystyle E_{N,Z_a}(t) = \frac{dZ_a(t)}{dt} - \beta Z(t) I(t) +
\mu_{Z_a} Z_a(t),
\end{array}
\end{equation}
and the plots of error functions are demonstrated in Fig. \ref{f3}
for $N=5,10$ and $\hbar=-0.8$.

\section{Conclusion}
The HATM is among of the accurate semi-analytical methods for
solving linear and non-linear problems based on its capabilities
such as operators, functions and parameters that we have freedom to
chose them. In this research, the HATM was applied to solve the
bio-mathematical model of HIV infection for CD8$^+$ T-cells.
Furthermore, the convergence theorem was proved that shows the
competency of HATM for solving non-linear problems. Based on the
numerical solutions for $N= 5, 10$ several $\hbar$-curves were
plotted that show the convergence regions of solutions. The
precision of method were demonstrated by plotting the residual error
functions.

~\\
~\\
~\\
~\\
~\\

\end{document}